\documentclass[preprintnumbers,amsmath,amssymb]{revtex4}
\usepackage{epsfig}
\usepackage{color}
\usepackage{dsfont}
\usepackage[colorlinks,urlcolor=blue,linkcolor=red,citecolor=blue]{hyperref}
\begin{document}
\setlength{\voffset}{1.0cm}
\title{Gross-Neveu model with O(2)$_L\times$O(2)$_R$ chiral symmetry:  Duality with
Zakharov-Mikhailov model and large $N$ solution}
\author{Michael Thies\footnote{michael.thies@gravity.fau.de}}
\affiliation{Institut f\"ur  Theoretische Physik, Universit\"at Erlangen-N\"urnberg, D-91058, Erlangen, Germany}
\date{\today}

\begin{abstract}
The two-flavor Gross-Neveu model with U(2)$_L\times$U(2)$_R$ chiral symmetry in 1+1 dimensions is used to construct 
a novel variant of four-fermion theories with O(2)$_L\times$O(2)$_R$ chiral symmetry. The spontaneous breaking of the group
O(2), a continuous group with two connected components (rotations and reflections), gives rise to new phenomena. It is ideally suited to describe
a situation where two distinct kinds of condensation compete, in particular chiral symmetry breaking (particle-hole condensation) and Cooper
pairing (particle-particle condensation). After solving the O(2) chiral Gross-Neveu model in detail, we demonstrate that it is dual to another classically 
integrable model due to Zakharov and Mikhailov. The duality enables us to solve the quantum version of this model in the large $N$ limit
with semiclassical methods, supporting its integrability at the quantum level. The resulting model is the unique four-fermion theory sharing the full
Pauli-G\"ursey symmetry with free, massless fermions (``perfect Gross-Neveu model") and provides us with a solvable model for competing chiral
and Cooper pair condensates, including explicit soliton dynamics and the phase diagram.
\end{abstract}

%\pacs{}
\maketitle

%<<<<<<<<<<<<<<<<<<<<<<<<<<<<<<<<<<<<<<<<<<<<<<<<<<<<<<<<<<<<<<<<<<<<<<<<<<<<<<<<<<<<<<<<<<<< <<<<<<<<<<<<<<<<<<<<<<<<<<<<<
%<<<<<<<<<<<<<<<<<<<<<<<<<<<<<<<<<<<<<<<<<<<<<<<<<<<<<<<<<<<<<<<<<<<<<<<<<<<<<<<<<<<<<<<<<<<<<<<<<<<<<<<<<<<<<<<<<<<<<<<<<<
\section{Introduction} 
\label{sect1}
%<<<<<<<<<<<<<<<<<<<<<<<<<<<<<<<<<<<<<<<<<<<<<<<<<<<<<<<<<<<<<<<<<<<<<<<<<<<<<<<<<<<<<<<<<<<<<<<<<<<<<<<<<<<<<<<<<<<<<<<<<<
%<<<<<<<<<<<<<<<<<<<<<<<<<<<<<<<<<<<<<<<<<<<<<<<

Back in 1978, Zakharov and Mikhailov \cite{L1} proved the integrability of three classical spinor models in 1+1 dimensions for any number of components $N$.
The quantum versions of two of them are by now also well under control, at least in the large $N$ limit, namely the Gross-Neveu (GN) model \cite{L2} 
\begin{equation}
{\cal L}_{\rm GN} = -2i \psi_1^{(i)*}\partial \psi_1^{(i)} + 2 i \psi_2^{(i)*} \bar{\partial} \psi_2^{(i)}  + \frac{g^2}{2} (  \psi_1^{(i)*} \psi_2^{(i)} + \psi_2^{(i)*} \psi_1^{(i)} )^2
\label{1.1}
\end{equation}
and the chiral GN model or two-dimensional (2D) Nambu--Jona-Lasinio (NJL) model \cite{L3}
\begin{equation}
{\cal L}_{\rm NJL} = -2i \psi_1^{(i)*}\partial \psi_1^{(i)} + 2 i \psi_2^{(i)*} \bar{\partial} \psi_2^{(i)} +2 g^2 ( \psi_1^{(i)*}\psi_2^{(i)}) ( \psi_2^{(j)*}\psi_1^{(j)}) .
\label{1.2}
\end{equation}
We use the notation  
\begin{equation}
z=x-t, \quad \bar{z}=x+t, \quad  \psi_1 = \psi_L ,  \quad \psi_2 = \psi_R
\label{1.3}
\end{equation}
and sum implicitly over ``color" indices $i,j$ from 1 to $N$.
Apparently, integrability at the classical level allows one to solve the quantized theory in the large $N$ limit with semiclassical methods, including time dependent
multisoliton interactions, in explicit analytical form \cite{L4,L5,L6}.
At the classical level (i.e., with $c$-number fermion fields), these two models are connected to chiral fields on the symplectic group Sp(2$N, \mathbb{R}$) (GN model)  or the special 
unitary group SU($N$) (NJL model). The third model presented in \cite{L1} and related to chiral fields on the orthogonal group O($N$) has so far not had any significant impact in particle physics.
It is sometimes referred to as Zakharov-Mikhailov (ZM) model and has the less familiar Lagrangian
\begin{equation}
{\cal L}_{\rm ZM} =   -2i \psi_1^{(i)*}\partial \psi_1^{(i)} + 2 i \psi_2^{(i)*} \bar{\partial} \psi_2^{(i)}   + g^2  ( \psi_1^{(i)*} \psi_1^{(j)} - \psi_1^{(j)*}\psi_1^{(i)})
( \psi_2^{(i)*} \psi_2^{(j)} - \psi_2^{(j)*}\psi_2^{(i)}).
\label{1.4}
\end{equation}
Interestingly, the quantum version of the ZM model has appeared again in a different context in the meantime.
When studying four-fermion theories that give rise to Cooper pairing as opposed to fermion-antifermion pairing, 
Chodos, Minakata and Cooper (CMC) \cite{L7} proposed a model whose Lagrangian is equivalent to
\begin{equation}
{\cal L}_{\rm CMC} =  -2i \psi_1^{(i)*}\partial \psi_1^{(i)} + 2 i \psi_2^{(i)*} \bar{\partial} \psi_2^{(i)} +2 g^2 ( \psi_1^{(i)}\psi_2^{(i)}) ( \psi_2^{(j)*}\psi_1^{(j)*}) .
\label{1.5}
\end{equation}
They noticed many similarities with the chiral GN model such as asymptotic freedom, mass generation and a massless bound state. If written in the form (\ref{1.2}) and  (\ref{1.5}),
one sees that the quantized NJL and CMC models
are ``dual" to each other in the sense that they are related by a simple Bogoliubov transformation \cite{L8}, 
\begin{equation}
\psi_1^{(i)} \to \psi_1^{(i)\dagger}, \quad \psi_2^{(i)} \to \psi_2^{(i)}.
\label{1.6}
\end{equation}
Hence both models are mathematically equivalent, although their physics looks quite different at first sight. This observation incited us to 
study yet another four-fermion theory obtained by ``self-dualizing" the NJL model, i.e. adding the interaction terms of models
(\ref{1.2}) and (\ref{1.5}) with the same coupling constant \cite{L9}. The resulting theory is singled out from all other variants of the GN model 
in that it shares the full Pauli-G\"ursey symmetry \cite{L10,L11} with free, massless Dirac fermions, in addition to a O($N$) color symmetry. 
Because of this high degree of symmetry, it has been dubbed ``perfect GN model" (pGN) in \cite{L12},
\begin{equation}
{\cal L}_{\rm pGN} = -2i \psi_1^{(i)*}\partial \psi_1^{(i)} + 2 i \psi_2^{(i)*} \bar{\partial} \psi_2^{(i)} +2 g^2 \left[ ( \psi_1^{(i)*}\psi_2^{(i)}) ( \psi_2^{(j)*}\psi_1^{(j)}) 
+ ( \psi_1^{(i)}\psi_2^{(i)}) ( \psi_2^{(j)*}\psi_1^{(j)*}) \right].
\label{1.7}
\end{equation}
As a matter of fact, the Lagrangians of the pGN model and the ZM model become identical once fermion fields are treated as anticommuting variables.
This observation has stimulated our interest in the large $N$ limit of the quantum ZM model (hereafter referred to as pGN model). However, a
systematic solution of the soliton problem or other questions has so far resisted all our attempts. Judging from the experience with the GN and NJL models, if the 
classical ZM model is integrable, one would expect the same kind of solvability for the GN, NJL and pGN models.

Here we propose a solution to this problem. We have found a duality between the {\em one-flavor} pGN model and the {\em two-flavor} NJL model, for which the general 
soliton solution has already been given at large $N$ \cite{L13,L14}.  In the case of the GN model, the most efficient way to solve soliton dynamics has been to 
start from the solution of the NJL model with U(1)$_L\times$U(1)$_R$ chiral symmetry and specialize to real mean field solutions,
thereby solving the O(1)$_L\times$O(1)$_R$ [or Z$_{2,L} \times$Z$_{2,R}$] GN model \cite{L6}. Here we generalize this approach by first
reducing known solutions of the U(2)$_L\times$U(2)$_R$ two-flavor NJL model to a novel O(2)$_L\times$O(2)$_R$ variant of the GN model, again
by selecting real mean fields. This model in turn will be shown to be dual to the pGN model, thus providing the key to the missing large $N$ solution of the pGN model.

This paper is organized as follows. After a reminder of some elementary facts about O(2) group theory in Sec.~\ref{sect2}, we propose the O(2)$_L\times$O(2)$_R$ symmetric
descendent of the unitary two-flavor chiral GN model in Sec.~{\ref{sect3}. The vacuum structure and gap equation are determined in Sec.~\ref{sect4}, the meson
spectrum in Sec.~\ref{sect5}, using the random phase approximation (RPA). Sections~\ref{sect6} and \ref{sect7} are dedicated to the most elementary solitonic
multifermion bound states, the kink, and interactions of several kinks. In Sec.~\ref{sect8} we then show that the analog of twisted kinks exist as 
constituents of bound states, similar to what happens in the GN model. The simplest breather is also constructed. Sec.~\ref{sect9} addresses a topic
well-known from the NJL model, namely massless multifermion bound states and inhomogeneous structures at finite chemical potentials (chiral spirals, 
kink-antikink crystal). Section~\ref{sect10} is perhaps the most important one of this paper. Here we show the equivalence between the O(2) chiral GN model
and the so far unsolved ZM (or pGN) model). In Sec.~\ref{sect11} we summarize our findings, reviewing the preceding results in the light of the duality
from a physical point of view.
  
%<<<<<<<<<<<<<<<<<<<<<<<<<<<<<<<<<<<<<<<<<<<<<<<<<<<<<<<<<<<<<<<<<<<<<<<<<<<<<<<<<<<<<<<<<<<< <<<<<<<<<<<<<<<<<<<<<<<<<<<<<
%<<<<<<<<<<<<<<<<<<<<<<<<<<<<<<<<<<<<<<<<<<<<<<<<<<<<<<<<<<<<<<<<<<<<<<<<<<<<<<<<<<<<<<<<<<<<<<<<<<<<<<<<<<<<<<<<<<<<<<<<<<
\section{Elementary group theory: from U(2) to O(2)} 
\label{sect2}
%<<<<<<<<<<<<<<<<<<<<<<<<<<<<<<<<<<<<<<<<<<<<<<<<<<<<<<<<<<<<<<<<<<<<<<<<<<<<<<<<<<<<<<<<<<<<<<<<<<<<<<<<<<<<<<<<<<<<<<<<<<
%<<<<<<<<<<<<<<<<<<<<<<<<<<<<<<<<<<<<<<<<<<<<<<<

In the defining representation, elements of the unitary group U(2)=U(1)$\times$SU(2) can be parametrized as
\begin{equation}
U = e^{-i\psi} e^{i \vec{n} \vec{\tau} \delta},\quad \vec{n} = \left( \begin{array}{c} \sin \theta \cos \phi \\ \sin \theta \sin \phi \\ \cos \theta \end{array} \right),
\label{2.1}
\end{equation}
with $\tau_i$ in the standard form of the Pauli matrices. An O(2) matrix is a real U(2) matrix. There are two distinct ways to get a real matrix out of (\ref{2.1}):
\begin{enumerate}

\item $ \theta = \phi = \pi/2, \psi=0 $ (hence $n_1=n_3=0$): 

\begin{equation}
R(\delta) = e^{ i\tau_2 \delta} =  \left( \begin{array}{rr}  \cos \delta &  \sin \delta \\ - \sin \delta & \cos \delta \end{array} \right), \quad {\rm det}R(\delta)=1.
\label{2.2}
\end{equation}

\item $ \psi = \delta = \pi/2, \phi=0 $ (hence $n_2=0$): 

\begin{eqnarray}
I(\theta) = n_1\tau_1+n_3\tau_3 & = &  \tau_3 e^{i \tau_2 \theta}  = 
   \left( \begin{array}{rr}  \cos \theta &  \sin \theta \\ \sin \theta & - \cos \theta \end{array} \right), \quad {\rm det} I(\theta) = -1.
\label{2.3}
\end{eqnarray}

\end{enumerate}
The matrix $R(\delta)$ corresponds to a rotation around the center of the ($x,y$) plane by an angle $\delta$. The matrix $I(\theta)$ is a rotation by an angle $\theta$ followed by a reflection
in the new $x$-axis. These two elements belong to the two connected components of the group O(2) characterized by their determinants $\pm 1$, with only rotations forming a subgroup, SO(2).
The group manifolds of U(2) and O(2) are S$_1\times$S$_3$ and S$_1$+S$_1$, respectively.
What we have done here is the two-dimensional analog of the transition from U(1) to O(1) (or Z$_2$) where one restricts the function $e^{-i\psi}$ to its real values, $\pm 1$.
Finally, note that O(2) is a non-Abelian group. Products of its elements can easily be evaluated 
with the help of 
\begin{equation}
I(\theta) = \tau_3 R(\theta) = R(-\theta) \tau_3, \quad R(\theta_1) R(\theta_2) = R(\theta_1+\theta_2).
\label{2.4}
\end{equation}

%<<<<<<<<<<<<<<<<<<<<<<<<<<<<<<<<<<<<<<<<<<<<<<<<<<<<<<<<<<<<<<<<<<<<<<<<<<<<<<<<<<<<<<<<<<<< <<<<<<<<<<<<<<<<<<<<<<<<<<<<<
%<<<<<<<<<<<<<<<<<<<<<<<<<<<<<<<<<<<<<<<<<<<<<<<<<<<<<<<<<<<<<<<<<<<<<<<<<<<<<<<<<<<<<<<<<<<<<<<<<<<<<<<<<<<<<<<<<<<<<<<<<<
\section{Gross-Neveu model with O(2)$_L\times$O(2)$_R$ symmetry} 
\label{sect3}
%<<<<<<<<<<<<<<<<<<<<<<<<<<<<<<<<<<<<<<<<<<<<<<<<<<<<<<<<<<<<<<<<<<<<<<<<<<<<<<<<<<<<<<<<<<<<<<<<<<<<<<<<<<<<<<<<<<<<<<<<<<
%<<<<<<<<<<<<<<<<<<<<<<<<<<<<<<<<<<<<<<<<<<<<<<<

The Lagrangian of the U(2)$_L \times$ U(2)$_R$ symmetric two-flavor NJL model reads \cite{L14}
\begin{equation}
{\cal L}_{\rm U(2)} = \bar{\psi} i \partial \!\!\!/ \psi + \frac{g^2}{4} \left[ (\bar{\psi}\psi)^2 + (\bar{\psi}\vec{\tau}\psi)^2 +  
(\bar{\psi} i \gamma_5 \psi)^2 + (\bar{\psi} i \gamma_5 \vec{\tau} \psi)^2 \right].
\label{3.1}
\end{equation}
We choose a ``chiral" representation of Dirac matrices ($\gamma_5$ diagonal),
\begin{equation}
\gamma^0 = \sigma^1, \quad \gamma^1 = i \sigma^2, \quad \gamma_5 = -\sigma_3.
\label{3.2}
\end{equation}
If we expand the isovector interaction terms, the interaction part of Lagrangian (\ref{3.1}) consists of eight squares of bilinears.
We can now generate simpler, Lorentz invariant Lagrangians by deleting some of these terms. We propose the following
choice, keeping only half of the interaction terms: 
\begin{equation}
{\cal L}_{\rm O(2)} =  \bar{\psi} i \partial \!\!\!/ \psi + \frac{g^2}{4} \left[ (\bar{\psi} \psi)^2 + (\bar{\psi}\tau_1  \psi)^2 + (\bar{\psi}\tau_3  \psi)^2+ (\bar{\psi}i \gamma_5 \tau_2  \psi)^2\right].
\label{3.3}
\end{equation}
Although it is not obvious, the resulting model has a O(2)$_L \times$O(2)$_R$ chiral symmetry. Let us evaluate the transformation of the 
four remaining bilinears induced by orthogonal chiral transformations of the spinor fields.
Isospin rotation of a left-handed spinor around the 2-axis,
\begin{equation}
P_L e^{i \alpha \tau_2} +P_R: \quad 
 \left( \begin{array}{c} \bar{\psi} \psi \\ \bar{\psi} i \gamma_5 \tau_2 \psi \end{array} \right)' =  R(-\alpha) 
\left( \begin{array}{c} \bar{\psi}\psi  \\ \bar{\psi}i \gamma_5 \tau_2 \psi  \end{array} \right), \quad \left( \begin{array}{c} \bar{\psi}\tau_3 \psi \\ \bar{\psi} \tau_1 \psi \end{array} \right)'= 
R(\alpha)  \left( \begin{array}{c} \bar{\psi}\tau_3 \psi \\ \bar{\psi} \tau_1 \psi \end{array}\right).
\label{3.4}
\end{equation}
Isospin rotation of a right-handed spinor around the 2-axis,
\begin{equation}
P_L+P_R e^{i\alpha \tau_2} : \quad
 \left( \begin{array}{c} \bar{\psi} \psi \\ \bar{\psi} i \gamma_5 \tau_2 \psi \end{array} \right)' = R(\alpha) 
\left( \begin{array}{c} \bar{\psi}\psi  \\ \bar{\psi}i \gamma_5 \tau_2 \psi  \end{array} \right), \quad \left( \begin{array}{c} \bar{\psi}\tau_3 \psi \\ \bar{\psi} \tau_1 \psi \end{array} \right)' =  
R(\alpha)  \left( \begin{array}{c} \bar{\psi}\tau_3 \psi \\ \bar{\psi} \tau_1 \psi \end{array}\right).
\label{3.5}
\end{equation}
Isospin reflection and rotation around the 2-axis of a left-handed spinor 
\begin{equation}
 P_L \tau_3 e^{i\alpha \tau_2} + P_R: \quad
 \left( \begin{array}{c} \bar{\psi} \psi \\ \bar{\psi} i \gamma_5 \tau_2 \psi \end{array} \right)' =   R(\alpha) 
\left( \begin{array}{c} \bar{\psi}\tau_3 \psi  \\ \bar{\psi}\tau_1 \psi  \end{array} \right), \quad \left( \begin{array}{c} \bar{\psi}\tau_3 \psi \\ \bar{\psi} \tau_1 \psi \end{array} \right)' =  
R(-\alpha) \left( \begin{array}{c} \bar{\psi}\psi \\ \bar{\psi} i \gamma_5 \tau_2 \psi \end{array}\right).
\label{3.6}
\end{equation}
Isospin reflection and rotation around the 2-axis of a right-handed spinor
\begin{equation}
P_L+ P_R \tau_3 e^{i \alpha \tau_2}: \quad
 \left( \begin{array}{c} \bar{\psi} \psi \\ \bar{\psi} i \gamma_5 \tau_2 \psi \end{array} \right)' = I(\alpha) 
\left( \begin{array}{c} \bar{\psi} \tau_3 \psi  \\ \bar{\psi} \tau_1 \psi  \end{array} \right), \quad \left( \begin{array}{c} \bar{\psi}\tau_3 \psi \\ \bar{\psi} \tau_1 \psi \end{array} \right)' =  
 I(\alpha) \left( \begin{array}{c} \bar{\psi}\psi \\ \bar{\psi} i \gamma_5 \tau_2 \psi \end{array}\right).
\label{3.7}
\end{equation}
Thus rotations leave the combinations $(\bar{\psi} \psi)^2 + (\bar{\psi}i\gamma_5 \tau_2 \psi)^2$ and $(\bar{\psi}\tau_1 \psi)^2 + (\bar{\psi}\tau_3 \psi )^2$ separately invariant.
Reflections leave only the sum of all four terms invariant, as they induce hopping between the two pairs of bilinears in addition to a rotation.

This confirms that model (\ref{3.3}) indeed possesses a O(2)$_L \times$O(2)$_R$ chiral symmetry. Consequently, the $\tau_2$ components of the isospin vector and axial vector currents
are conserved,
\begin{equation}
\partial_{\mu} \bar{\psi} \gamma^{\mu} \tau_2 \psi = 0, \quad \partial_{\mu} \bar{\psi} \gamma^{\mu} \gamma_5 \tau_2 \psi = 0.
\label{3.8}
\end{equation}
In addition the model has SU($N$) color symmetry and U(1) fermion number ($\psi \to e^{i\alpha} \psi$). It also shares with the GN model charge conjugation
with the familiar consequences, i.e., a real mean field in the Hartree-Fock (HF) approach and a fermion spectrum that is symmetric about 0,
\begin{equation}
C: \quad \psi \to \gamma_5 \psi^*, \quad \gamma_5 H^* \gamma_5 = - H.
\label{3.9}
\end{equation}
Here, $H$ is the HF Hamiltonian in coordinate space. In view of the conserved charges, there are three chemical potentials one can add to the HF Hamiltonian,
\begin{equation}
{\cal H} \to {\cal H} - \mu \psi^{\dagger}\psi - \mu_2 \psi^{\dagger} \tau_2 \psi - \mu_{5,2} \psi^{\dagger} \gamma_5 \tau_2 \psi.
\label{3.10}
\end{equation}
This is important if one considers the phase diagram of the model.

%<<<<<<<<<<<<<<<<<<<<<<<<<<<<<<<<<<<<<<<<<<<<<<<<<<<<<<<<<<<<<<<<<<<<<<<<<<<<<<<<<<<<<<<<<<<< <<<<<<<<<<<<<<<<<<<<<<<<<<<<<
%<<<<<<<<<<<<<<<<<<<<<<<<<<<<<<<<<<<<<<<<<<<<<<<<<<<<<<<<<<<<<<<<<<<<<<<<<<<<<<<<<<<<<<<<<<<<<<<<<<<<<<<<<<<<<<<<<<<<<<<<<<
\section{Vacua and dynamical mass} 
\label{sect4}
%<<<<<<<<<<<<<<<<<<<<<<<<<<<<<<<<<<<<<<<<<<<<<<<<<<<<<<<<<<<<<<<<<<<<<<<<<<<<<<<<<<<<<<<<<<<<<<<<<<<<<<<<<<<<<<<<<<<<<<<<<<
%<<<<<<<<<<<<<<<<<<<<<<<<<<<<<<<<<<<<<<<<<<<<<<<

We start solving the chiral O(2) GN model (\ref{3.3}) in the large $N$ limit by determining its gap equation and vacuum structure. The HF equation reads
\begin{equation}
(i \gamma^{\mu}\partial_{\mu} -S_0-S_1\tau_1-S_3\tau_3 - i \gamma_5 P_2 \tau_2)\psi  = 0
\label{4.1}
\end{equation}
with the mean fields given by the vacuum expectation values
\begin{equation}
\left( \begin{array}{c} S_0 \\ S_1 \\ S_3 \\ P_2 \end{array} \right) = - \frac{g^2}{2} \left( \begin{array}{c} \langle \bar{\psi}\psi \rangle \\
\langle \bar{\psi} \tau_1 \psi \rangle \\ \langle \bar{\psi} \tau_3 \psi \rangle \\ \langle \bar{\psi}i \gamma_5  \tau_2 \psi \rangle \end{array} \right) .
\label{4.2}
\end{equation}
Denoting two-component isospinors of chirality $L/R$ by $\psi_{1,2}$, Eq.~(\ref{4.1}) may be rewritten in canonical form as
\begin{equation}
i \partial_t \left( \begin{array}{c} \psi_1 \\ \psi_2 \end{array} \right) = \left( \begin{array}{cc} i \partial_x & \Delta^T \\ \Delta & -i \partial_x \end{array} \right)
\left( \begin{array}{c} \psi_1 \\ \psi_2 \end{array} \right).
\label{4.3}
\end{equation}
Here, $\Delta$ is the 2$\times$2 flavor matrix 
\begin{equation}
\Delta = S_0 + S_1 \tau_1 + S_3 \tau_3 -i P_2 \tau_2 = \left( \begin{array}{cc} S_0+S_3 & S_1-P_2 \\ S_1+P_2 & S_0-S_3 \end{array} \right).
\label{4.4}
\end{equation}
The fact that this matrix is real reflects the orthogonal symmetry.
To determine the HF ground state, we have to diagonalize the 4$\times$4 Hamiltonian 
\begin{equation}
h = \left( \begin{array}{cc} -k & \Delta^T \\ \Delta & k \end{array} \right)
\label{4.5}
\end{equation}
with a constant, real matrix $\Delta$. We find four modes with energies $\pm\sqrt{m_1^2+k^2}, \pm\sqrt{m_2^2+k^2}$ where 
\begin{equation}
m_{1,2}^2 = \left( \sqrt{S_0^2+P_2^2} \pm \sqrt{S_1^2+ S_3^2} \right)^2.
\label{4.6}
\end{equation}
Minimizing the HF vacuum energy density
\begin{equation}
{\cal E}_{\rm vac} = - N \int_{-\Lambda/2}^{\Lambda/2} \frac{dk}{2\pi} \left( \sqrt{k^2+m_1^2} + \sqrt{k^2+ m_2^2}\, \right) + \frac{m_1^2+m_2^2}{2 g^2}
\label{4.7}
\end{equation}
with respect to $m_1,m_2$ yields two gap equations
\begin{equation}
0 = 1 + \frac{N g^2}{2\pi} \ln \frac{m_i^2}{\Lambda^2} \quad (i=1,2)
\label{4.8}
\end{equation}
that are compatible only if $m_1=m_2=m$. The final gap equation is identical to that of the GN model and has the same implications
(dimensional transmutation, asymptotic freedom). The renormalized vacuum energy density becomes
\begin{equation}
{\cal E}_{\rm vac} = - N \frac{m^2}{2\pi}.
\label{4.9}
\end{equation}
According to Eq.~(\ref{4.6}), there are two possible choices leading to $m_1=m_2$: 
\begin{eqnarray}
1) & & S_1=S_3=0, \quad S_0^2 + P_2^2=m^2, \quad S_0=m \cos \alpha, \quad P_2 = m \sin \alpha, \quad \Delta =  m R(-\alpha)
\nonumber \\
2) & & S_0=P_2=0, \quad S_1^2 + S_3^2=m^2, \quad S_3=m \cos \alpha, \quad S_1 = m \sin \alpha, \quad \Delta = m I(\alpha) 
\label{4.10}
\end{eqnarray}
As expected, the vacuum manifold is the group manifold of O(2) comprising two disjoint circles in the ($S_0,P_2$) and ($S_1,S_3)$ planes.
For simplicity we shall denote a vacuum in the ($S_0,P_2$) plane as ``rotation vacuum" and in the
($S_1,S_3$) plane as ``reflection vacuum" to indicate their origin in the two components of the group O(2). When breaking spontaneously the
O(2) chiral symmetry, the system has to pick one circle (rotation or reflection) and a particular point on this circle (angle $\alpha$).
All of these vacua are, of course, on an equal footing. If not indicated otherwise, our standard choice will be ($S_0=m,P_2=0, \Delta=m$)
for the rotation vacuum and  ($S_1=0,S_3=m,\Delta=m \tau_3$) for the reflection vacuum, so that $\Delta$ is diagonal in the standard vacua.

%<<<<<<<<<<<<<<<<<<<<<<<<<<<<<<<<<<<<<<<<<<<<<<<<<<<<<<<<<<<<<<<<<<<<<<<<<<<<<<<<<<<<<<<<<<<< <<<<<<<<<<<<<<<<<<<<<<<<<<<<<
%<<<<<<<<<<<<<<<<<<<<<<<<<<<<<<<<<<<<<<<<<<<<<<<<<<<<<<<<<<<<<<<<<<<<<<<<<<<<<<<<<<<<<<<<<<<<<<<<<<<<<<<<<<<<<<<<<<<<<<<<<<
\section{Meson spectrum in random phase approximation} 
\label{sect5}
%<<<<<<<<<<<<<<<<<<<<<<<<<<<<<<<<<<<<<<<<<<<<<<<<<<<<<<<<<<<<<<<<<<<<<<<<<<<<<<<<<<<<<<<<<<<<<<<<<<<<<<<<<<<<<<<<<<<<<<<<<<
%<<<<<<<<<<<<<<<<<<<<<<<<<<<<<<<<<<<<<<<<<<<<<<<

Here we follow closely previous works on other GN model variants, using the equations of motion method for the one-body density matrix $Q(x,y)$ \cite{L15,L16}.
The starting point is the equation
\begin{eqnarray}
i \partial_t Q(x,y) & = & -i \left[ \partial_y Q(x,y) \gamma_5 + \gamma_5 \partial_x Q(x,y) \right]
\nonumber \\
& & - \frac{N g^2}{2} \sum_{n=1}^4 \left\{ {\rm Tr} [{\cal O}_n Q(x,y) ] {\cal O}_n Q(x,y) -Q(x,y) {\cal O}_n {\rm Tr}[{\cal O}_n Q(x,y)] \right\}
\label{5.1}
\end{eqnarray}
with
\begin{equation}
{\cal O}_1 = \gamma^0, \quad {\cal O}_2 = \gamma^0 \tau_1, \quad {\cal O}_3 = \gamma^0 \tau_3 , \quad {\cal O}_4 = i \gamma^1 \tau_2
\label{5.2}
\end{equation}
matching the interactions in Lagrangian (\ref{3.3}). Let us first assume the rotation vacuum ($S_0=m,P_2=0$).
Using free positive and negative energy spinors for one flavor ($E_k=\sqrt{k^2+m^2}$)
\begin{equation}
\left( \begin{array}{cc} -k & m \\ m & k \end{array} \right) u(k) = E_k u(k), \quad u(k)= \left( \begin{array}{c} u_1 \\ u_2 \end{array} \right) = 
\frac{1}{\sqrt{2 E_k(E_k+k)}} \left( \begin{array}{c} m \\ E_k+k \end{array} \right),
\label{5.3}
\end{equation}
\begin{equation}
\left( \begin{array}{cc} -k & m \\ m & k \end{array} \right) v(k) = - E_k v(k), \quad v(k)= \left( \begin{array}{c} v_1 \\ v_2 \end{array} \right) = 
\frac{1}{\sqrt{2 E_k(E_k-k)}} \left( \begin{array}{c} m \\ k - E_k \end{array} \right),
\label{5.4}
\end{equation}
the corresponding free massive states in the two-flavor case are
\begin{equation}
u_{I} = \left( \begin{array}{c} u_1 \\ 0 \\ u_2 \\ 0 \end{array} \right), \quad 
u_{II} = \left( \begin{array}{c} 0 \\ u_1 \\ 0 \\ u_2  \end{array} \right), \quad 
v_{I} = \left( \begin{array}{c} v_1 \\ 0 \\ v_2 \\ 0 \end{array} \right), \quad 
v_{II} = \left( \begin{array}{c} 0 \\ v_1 \\ 0 \\ v_2  \end{array} \right) .
\label{5.5}
\end{equation}
The isospin labels $I,II$ will be denoted by Greek letters below.
Linearizing Eq.~(\ref{5.1}) in the fluctuation around the vacuum density matrix and sandwiching it between vacuum and one meson states of momentum $P$ and energy 
${\cal E}(P)$, we arrive at the RPA equations 
\begin{eqnarray}
{\cal E}(P) X_{\alpha \beta}(P,k) & = &  E(k-P,k) X_{\alpha \beta}(P,k) - \frac{Ng^2}{2} \sum_{n=1}^4 v_{\alpha}^{\dagger}(k-P) {\cal O}_n u_{\beta}(k) \Phi_n(P),
\nonumber \\
{\cal E}(P) Y_{\alpha \beta}(P,k) & = &  -E(k-P,k) Y_{\alpha \beta}(P,k) +  \frac{Ng^2}{2} \sum_{n=1}^4  u_{\alpha}^{\dagger}(k-P) {\cal O}_n v_{\beta}(k) \Phi_n(P),
\nonumber \\
E(k-P,k) & = & E_{k-P} + E_k,
\label{5.6}
\end{eqnarray}
with
\begin{equation}
\Phi_n(P)  =   \int \frac{dq}{2\pi} \left\{ {\cal B}^n_{\delta \gamma}(P,q) Y_{\gamma \delta}(P,q) 
+ {\cal C}^n_{\delta \gamma}(P,q)  X_{\gamma \delta}(P,q) \right\}
\label{5.7}
\end{equation}
and
\begin{eqnarray}
{\cal B}^n_{\delta \gamma}(P,q) & = & v_{\delta}^{\dagger}(q) {\cal O}_n u_{\gamma}(q-P),
\nonumber \\
{\cal C}^n_{\delta \gamma}(P,q) & = & u_{\delta}^{\dagger}(q) {\cal O}_n v_{\gamma}(q-P).
\label{5.8}
\end{eqnarray}
Equations~(\ref{5.6}) are integral equations with a separable kernel as is characteristic for all GN-type models. They can thus be solved analytically.
By first solving Eq.~(\ref{5.6}) for $X_{\alpha \beta}$ and $Y_{\alpha \beta}$, we get
\begin{eqnarray}
X_{\alpha \beta}(P,k) & = & \sum_{n=1}^4 {\cal A}_{\alpha \beta}^n (P,k) \Phi_n(P) ,
\nonumber \\
Y_{\alpha \beta}(P,k) & = &  \sum_{n=1}^4 {\cal D}_{\alpha \beta}^n (P,k) \Phi_n(P), 
\label{5.9}
\end{eqnarray}
with the definitions
\begin{eqnarray}
{\cal A}_{\alpha \beta}^n (P,k) & = & -Ng^2 \frac{1}{{\cal E}(P) - E(k-P,k)} v_{\alpha}^{\dagger}(k-P) {\cal O}_n u_{\beta}(k),
\nonumber \\
{\cal D}_{\alpha \beta}^n (P,k) & = & Ng^2 \frac{1}{{\cal E}(P) + E(k-P,k)} u_{\alpha}^{\dagger}(k-P) {\cal O}_n v_{\beta}(k).
\label{5.10}
\end{eqnarray}
Inserting the results (\ref{5.9}) into Eq.~(\ref{5.7}) leads to the homogeneous linear system for $\Phi_n(P)$,
\begin{equation}
\Phi_n(P) = \sum_m {\cal M}_{nm} (P) \Phi_m(P).
\label{5.11}
\end{equation}
The 4$\times$4 matrix ${\cal M}_{nm}(P)$  is given by
\begin{equation}
{\cal M}_{nm}(P) = \int \frac{dq}{2\pi} \left\{ {\cal B}_{\delta \gamma}^n(P,q) {\cal D}_{\gamma \delta}^m(P,q)
+  {\cal C}_{\delta \gamma}^n(P,q) {\cal A}_{\gamma \delta}^m(P,q) \right\}.
\label{5.12}
\end{equation}
An explicit analytical calculation shows that ${\cal M}_{nm}$ is diagonal, hence Eq.~(\ref{5.11}) 
reduces to
\begin{equation}
{\cal M}_{nn} = 1 \quad ({\rm no\ sum}).
\label{5.14}
\end{equation}
One finds only two different diagonal matrix elements ${\cal M}_{nn}$. In the isovector pseudoscalar channel $P_2$,
\begin{equation}
{\cal M}_{44} = \frac{Ng^2}{2} \int \frac{dk}{2\pi} \left( \frac{1}{E(k-P)} + \frac{1}{E(k)} \right) \left( \frac{P^2-E^2(k-P,k)}{{\cal E}^2(P) -E^2(k-P,k)} \right).
\label{5.15}
\end{equation}
The choice ${\cal E}^2(P)=P^2$ converts the condition ${\cal M}_{44}=1$ into the vacuum gap equation. 
This proves the existence of a massless mode, the ``would be Goldstone boson" fluctuating in the direction tangential to the vacuum circle.
In the other three channels, there is a (marginally bound) massive state with the common mass $2m$, the same as in the GN model.
The corresponding diagonal matrix elements ${\cal M}_{nn}$ read
\begin{equation}
{\cal M}_{11}={\cal M}_{22} = {\cal M}_{33} = \frac{Ng^2}{2} \int \frac{dk}{2\pi} \left( \frac{1}{E(k-P)} + \frac{1}{E(k)} \right) \left( \frac{4m^2+P^2-E^2(k-P,k)}{{\cal E}^2(P) -E^2(k-P,k)} \right).
\label{5.16}
\end{equation}
Here the ansatz ${\cal E}^2(P)=4m^2+P^2$ yields again back to the gap equation. The massive bound states are scalar mesons in the $S_0,S_1,S_3$ channels.

One can repeat the same calculation by starting from the  reflection vacuum ($S_3=m,S_1=0$). The only change in the formalism is the fact that 
the spinors for isospin down have to be evaluated with $m \to -m$, since the vacuum has $\Delta=m \tau_3$ rather than $\Delta=m$.
The results are the same, except that the massless mode now appears in the $S_1$ channel, again tangential to the reflection vacuum circle
in the chosen vacuum point. The mesons in the $S_0,S_3,P_2$ channels all have the same mass 2$m$. At first sight, it looks as if now
a scalar meson would be massless and the pseudoscalar one massive. This is not the case. When using the rotation vacuum, the parity 
operation on spinors has the usual form,
\begin{equation}
P: \quad \psi(x) \to \gamma^0 \psi(-x),
\label{5.17}
\end{equation}
since the vacuum is also standard. To reach the reflection vacuum $m \tau_3$, one has to perform the chiral transformation
$P_L \tau_3 + P_R$. Under this transformation, the matrix $\gamma^0$ in (\ref{5.17}) goes over into $\tau_3 \gamma^0$, so that now $S_0,S_3,P_2$ are scalars
whereas $S_1$ is pseudoscalar.
 
It is noteworthy that all massive mesons found in GN-type models so far share the common mass $M=2m$.
This does not mean that they are noninteracting, otherwise the dispersion relation would not be ${\cal E}(P)=\sqrt{4 m^2+ P^2}$.
This ``universality" is most likely a side effect of the large $N$ limit. The RPA goes beyond the leading order (HF) by taking into account fluctuations
of O($1/\sqrt{N}$). It is plausible that meson binding energies are suppressed by at least a factor of $1/N$ and therefore not yet visible at this order
of the large $N$ expansion.

We conclude this chapter with a table comparing the meson content of various chirally symmetric GN-type models, to emphasize the common 
aspects as well as the differences due to different symmetries. Only two rules govern the whole picture in Table \ref{tab1}: The total number of mesons is equal
to the number of squares of bilinears in the interaction Lagrangian, and the number of massless modes equals the number of flat directions on
the vacuum manifold.

\begin{center}
\begin{table}
\begin{tabular}{|c|c|c|c|c|c|}
\hline
   & Chiral symmetry &  Vacuum manifold &  Mesons  &  Massless  & Massive    \\
\hline
 GN & O(1)$_L\times$O(1)$_R$ & Z$_2$ & 1 & 0 & 1 \\
 NJL & U(1)$_L\times$U(1)$_R$ & S$_1$ & 2 & 1 & 1 \\
 isoNJL & SU(2)$_L \times$SU(2)$_R$ & S$_3$ & 4 & 3 & 1 \\
O(2) GN & O(2)$_L \times$O(2)$_R$ & S$_1+$S$_1$ & 4 & 1 & 3 \\
U(2) GN & U(2)$_L \times$U(2)$_R$ & S$_1 \times$S$_3$ & 8 & 4 & 4 \\
\hline
\end{tabular}
\caption{Overview of meson content of different chiral GN models.}
\label{tab1}
\end{table}
\end{center}

%<<<<<<<<<<<<<<<<<<<<<<<<<<<<<<<<<<<<<<<<<<<<<<<<<<<<<<<<<<<<<<<<<<<<<<<<<<<<<<<<<<<<<<<<<<<< <<<<<<<<<<<<<<<<<<<<<<<<<<<<<
%<<<<<<<<<<<<<<<<<<<<<<<<<<<<<<<<<<<<<<<<<<<<<<<<<<<<<<<<<<<<<<<<<<<<<<<<<<<<<<<<<<<<<<<<<<<<<<<<<<<<<<<<<<<<<<<<<<<<<<<<<<
\section{Basic kink} 
\label{sect6}
%<<<<<<<<<<<<<<<<<<<<<<<<<<<<<<<<<<<<<<<<<<<<<<<<<<<<<<<<<<<<<<<<<<<<<<<<<<<<<<<<<<<<<<<<<<<<<<<<<<<<<<<<<<<<<<<<<<<<<<<<<<
%<<<<<<<<<<<<<<<<<<<<<<<<<<<<<<<<<<<<<<<<<<<<<<<

Kink denotes a multifermion bound state connecting two different vacua at $x \to \pm \infty$. We use units where $m=1$ from now on to make contact with  
the literature. To set the stage, let us briefly recall the twisted kink of the one-flavor NJL model
originally due to Shei \cite{L17}. The kink at rest connecting the U(1) vacua $e^{i \alpha}$ at $x\to - \infty$ and $e^{i \beta}$ at $x\to \infty$ has the mean field
\begin{equation}
\Delta(x) = e^{i \alpha} (1-f(x)) + e^{i\beta} f(x), \quad f(x) = \frac{V(x)}{1+ V(x)} , \quad V(x) = e^{2x \sin \theta}.
\label{6.1}
\end{equation}
We have written it in a form where the interpolating structure is clear, since $f$ and $(1-f)$ are kinklike scalar functions. The angle $\theta$ in $V(x)$
is called twist angle and related to the difference of the two asymptotic vacuum phases via
\begin{equation}
\theta = \frac{1}{2} (\alpha- \beta).
\label{6.2}
\end{equation}
Since U(1) is Abelian and $\Delta$ transforms under chiral transformations as follows
\begin{equation}
\psi_1 \to e^{i \alpha_1} \psi_1, \quad \psi_2 \to e^{i \alpha_2} \psi_2, \quad \Delta \to e^{i \alpha_1} \Delta e^{-i \alpha_2},
\label{6.3}
\end{equation}
the twist angle is chirally invariant. As such it has a physical meaning, determining the slope of the kink
profile and its fermion number. There is a bound state with energy $\epsilon_0=\cos \theta$, occupied by $N \sin \theta$ fermions.
If we restrict ourselves to real $\Delta$ as appropriate for the GN model with discrete chiral symmetry, we have to choose $\theta=\pm\pi/2$ and either $\alpha=\pi, \beta=0$
(kink) or $\alpha=0, \beta=\pi$ (antikink). Here, the kink profile can only assume the steepest shape. The bound state moves to 0 energy, the center of the mass gap.
We get
\begin{equation}
\Delta(x) \to S(x) = \pm \tanh x,
\label{6.4}
\end{equation} 
an early result attributed to Callan, Coleman, Gross and Zee in Refs.~\cite{L18,L19}. 

Let us repeat the reduction from unitary to orthogonal chiral symmetry, now for two flavors. We can start directly from the known result for the twisted kink in the
U(2)$_L \times$U(2)$_R$ GN model \cite{L13,L14}. If the vacuum at $x\to - \infty$ is taken as 1, the formalism yields the expression
\begin{equation}
\Delta(x) = \frac{1+U V(x)}{1+V(x)}
\label{6.5}
\end{equation} 
with 
\begin{equation}
V(x)  =  e^{2 \sin \theta x}, \quad 
U  =  (1 - \vec{p}\, \vec{p}^{\,\dagger})+ e^{-2i\theta} \vec{p}\, \vec{p}^{\,\dagger} .
\label{6.6}
\end{equation}
$U$ is the vacuum at $x \to \infty$, the vector $\vec{p}$ a complex, two-dimensional vector normalized to $\vec{p}^{\,\dagger} \vec{p}=1$.
We can interpret the expression for $U$ as spectral representation of a unitary 2$\times$2 matrix with eigenvalues 1 and $e^{-2i\theta}$.
Choosing a frame where $\vec{p}=(1,0)$, we recover the NJL twisted kink for isospin up and the vacuum for isospin down.
This reduces the U(2) twisted kink to the U(1) twisted kink, with $\theta$ the (chirally invariant) twisting angle. The unique way to
get a real solution as needed for the O(2) model is again to choose the maximal twist angle, $\theta=\pi/2$. We then recover the 
real GN kink in the isospin up state. From the O(2) point of view, the kink connects the rotational vacuum 1 with the reflection vacuum $\tau_3$.
It is impossible to find a kink connecting two points on the same vacuum circle. Going back to the form (\ref{6.6}) of $U$, we can construct a more
general kink by choosing a real $\vec{p}$,
\begin{equation}
\vec{p} = \left( \begin{array}{c} \cos \alpha \\ \sin \alpha \end{array} \right).
\label{6.7}
\end{equation}
The mean field then becomes the matrix
\begin{equation}
\Delta = (1-f) - I(2 \alpha) f, \quad f= \frac{V}{1+V} = \frac{1}{2} (1+ \tanh x).
\label{6.8}
\end{equation}
By a further chiral transformation we arrive at the most general O(2) kink interpolating between arbitrary points on the rotation and reflection circles,
\begin{equation}
\Delta = R(\alpha_1) (1-f) -I(\alpha_2) f.
\label{6.9}
\end{equation}
Here one has to choose $\alpha= \alpha_1+\alpha_2$ in $\vec{p}$. This kink evolves from the rotation vacuum $R(\alpha_1)$ at $x\to - \infty$ 
to the reflection vacuum $-I(\alpha_2)$ at $x\to \infty$. Unlike the twist angle in the unitary models, here the angles 
$\alpha_1,\alpha_2$ have no direct physical significance, being dependent on the chiral frame chosen. This is consistent with the fact 
that the $x$ dependence is independent of the angles $\alpha_i$. 
The true twist angle $\theta=\pi/2$ is always maximal, like in the GN model.
The opposite kink connecting a point on the reflection circle to one on the rotation circle can easily be found by changing the sign of $x$,
\begin{equation}
\Delta = -I(2 \alpha_2) (1-f) + R(2\alpha_1) f.
\label{6.10}
\end{equation}
The reason why we have written down the most general form of the kinks is the fact that in scattering or bound state configurations with several kinks,  
the vectors $\vec{p}_i$ cannot all be rotated simultaneously into a specific direction. Then the differences $\alpha_i-\alpha_j$ do acquire physical
significance.

Finally we mention that the mass of the kink is the same as in the GN model, $M=N/(2\pi)$, independently of its fermion number.

%<<<<<<<<<<<<<<<<<<<<<<<<<<<<<<<<<<<<<<<<<<<<<<<<<<<<<<<<<<<<<<<<<<<<<<<<<<<<<<<<<<<<<<<<<<<< <<<<<<<<<<<<<<<<<<<<<<<<<<<<<
%<<<<<<<<<<<<<<<<<<<<<<<<<<<<<<<<<<<<<<<<<<<<<<<<<<<<<<<<<<<<<<<<<<<<<<<<<<<<<<<<<<<<<<<<<<<<<<<<<<<<<<<<<<<<<<<<<<<<<<<<<<
\section{Scattering of kinks} 
\label{sect7}
%<<<<<<<<<<<<<<<<<<<<<<<<<<<<<<<<<<<<<<<<<<<<<<<<<<<<<<<<<<<<<<<<<<<<<<<<<<<<<<<<<<<<<<<<<<<<<<<<<<<<<<<<<<<<<<<<<<<<<<<<<<
%<<<<<<<<<<<<<<<<<<<<<<<<<<<<<<<<<<<<<<<<<<<<<<<

We first use the existing tools to derive scattering of two twisted kinks in the U(2)$_L \times$U(2)$_R$ GN model \cite{L14}. In order to turn the result into a
solution of the O(2)$_L \times$O(2)$_R$ symmetric model, we choose as input two real vectors 
\begin{equation}
\vec{p}_i = \left( \begin{array}{c} \cos \alpha_i \\ \sin \alpha_i \end{array} \right),
\label{7.1}
\end{equation}
twist angles $\theta_i = \pi/2$ and positions of poles in the complex spectral parameter plane 
\begin{equation}
\zeta_i = \frac{i}{\eta_i}, \quad \eta_i = \sqrt{\frac{1+v_i}{1-v_i}}.
\label{7.2}
\end{equation}
We also introduce the ratio $\eta=\eta_1/\eta_2$.
The matrix $\omega$ is taken to be diagonal.
The result for kink-antikink scattering in the O(2) model are the real mean fields 
\begin{eqnarray}
{\cal D} S_0 & = & 1- \left(1-\frac{2(1+\eta^2)\cos^2 \alpha_{12}}{(1+\eta)^2} \right) V_1 V_2,
\nonumber \\
{\cal D} S_1 & = & - V_1 \sin 2\alpha_1 - V_2 \sin 2\alpha_2,
\nonumber \\
{\cal D} S_3 & = &  - V_1 \cos 2\alpha_1 - V_2 \cos 2\alpha_2,
\nonumber \\
{\cal D} P_2 & = & - \left( \frac{1-\eta}{1+\eta} \right) \sin2 \alpha_{12} V_1 V_2,
\label{7.3}
\end{eqnarray}
with $\alpha_{12}=\alpha_1- \alpha_2$. Here, ${\cal D}$ is the common denominator
\begin{equation}
{\cal D} =  1 + V_1 + V_2 + \kappa V_1 V_2, \quad \kappa= \left( 1- \frac{4 \eta \cos^2 \alpha_{12}}{(1+\eta)^2} \right) .
\label{7.4}
\end{equation}
The $V_i$ factors carry the ($x,t$) dependence, 
\begin{equation}
V_i = e^{2 x_i'}, \quad x_i'=\frac{x-x^0_i - v_it}{\sqrt{1-v_i^2}}.
\label{7.5}
\end{equation}
The vacuum at $x \to - \infty$ has been taken to be 1. Orthogonal matrices $U_1,U_2$ appear in intermediate states of the scattering process, whereas $U_{12}$ is the vacuum at $x\to \infty$.
One finds
\begin{equation}
U_i = 1-2 \vec{p}_i\vec{p}_i^{\,\dagger} = -I(2\alpha_i) 
\label{7.6}
\end{equation}
and
\begin{equation}
U_{12} = \frac{1}{1+\eta^2- 2 \eta \cos 2\alpha_{12}} \left( \begin{array}{cc} (1+\eta^2)\cos 2 \alpha_{12}  - 2 \eta  &
(1-\eta^2) \sin 2 \alpha_{12} \\ - (1-\eta^2) \sin 2 \alpha_{12} & (1+\eta^2)\cos 2 \alpha_{12} - 2 \eta  \end{array} \right).
\label{7.7}
\end{equation}
$U_{12}$ is a rotation matrix, in contrast to the reflection matrices $U_i$,
\begin{equation}
U_{12} = R(\Phi), \quad \tan \Phi = \frac{(1-\eta^2) \sin 2 \alpha_{12}}{ (1+\eta^2)\cos 2 \alpha_{12} - 2 \eta }.
\label{7.8}
\end{equation}
The mean field for the kink-antikink collision can be concisely represented as
\begin{equation}
\Delta = \frac{1 + V_1 U_1 + V_2 U_2 + \kappa V_1 V_2 U_{12}}{1+V_1+V_2+\kappa V_1 V_2} 
\label{7.9}
\end{equation}
where the four matrices ($1,\, U_1,\,U_2,\,U_{12}$) are now orthogonal as opposed to unitary in the U(2) model.
The physical meaning of these matrices is that they represent all possible vacua if the kinks are well separated in space \cite{L14}.

Let us pause for a moment and ask ourselves what is the intrinsic (chiral frame independent) content of expression (\ref{7.9}).
Chiral symmetry has already been employed to set the initial vacuum equal to 1. The only residual O(2) transformations allowed
without changing the initial vacuum are
\begin{equation}
\Delta \to R(\beta) \Delta R(-\beta), \quad \Delta \to I(\gamma) \Delta I(\gamma).
\label{7.10}
\end{equation}
If applied to the final vacuum $U_{12}$, the rotation acts as the identity whereas the reflection changes the sign of $\Phi$, or, equivalently, interchanges
$\vec{p}_1$ and $\vec{p}_2$. In the intermediate vacua $U_1,U_2$, both angles are shifted by the same amount. This reflects the fact that the only
chirally invariant quantity one can form out of the $\vec{p}_i$ is the scalar product $\vec{p}_1 \vec{p}_2$. 

Due to the four components $S_0,S_1,S_3,P_2$, the kink-kink collision process looks rather complicated. What happens can be described qualitatively 
as follows. Let us assume that kink 1 is incident from the left, kink 2 from the right. The asymptotic vacua are the rotational vacua 1 (at $x \to - \infty$)
and $R(\Phi)$ (at $x \to \infty$) throughout the collision process. In between the kinks, the system is in the reflection vacuum $U_1$ before the collision
and $U_2$ after the collision. The region where the system is in the reflection vacuum is always bounded by the position of the two kinks. This is
exactly what one would expect from a collision between two domain walls, here separating the rotational from the reflection phases.
The difference between the U(2) and the O(2) chirally symmetric cases is the fact that the intrinsic twist of the kinks is always maximal 
in the O(2) model, similar to the difference between the U(1) (NJL) and O(1) (GN) models.

Finally, let us remark that one also finds a bound state at rest by specializing to $\eta_1=\eta_2=1$. This is similar to the unitary models, but no such bound state 
exists in the one-flavor GN model.

\newpage
%<<<<<<<<<<<<<<<<<<<<<<<<<<<<<<<<<<<<<<<<<<<<<<<<<<<<<<<<<<<<<<<<<<<<<<<<<<<<<<<<<<<<<<<<<<<< <<<<<<<<<<<<<<<<<<<<<<<<<<<<<
%<<<<<<<<<<<<<<<<<<<<<<<<<<<<<<<<<<<<<<<<<<<<<<<<<<<<<<<<<<<<<<<<<<<<<<<<<<<<<<<<<<<<<<<<<<<<<<<<<<<<<<<<<<<<<<<<<<<<<<<<<<
\section{Baryon and breather with twisted kink constituents} 
\label{sect8}
%<<<<<<<<<<<<<<<<<<<<<<<<<<<<<<<<<<<<<<<<<<<<<<<<<<<<<<<<<<<<<<<<<<<<<<<<<<<<<<<<<<<<<<<<<<<<<<<<<<<<<<<<<<<<<<<<<<<<<<<<<<
%<<<<<<<<<<<<<<<<<<<<<<<<<<<<<<<<<<<<<<<<<<<<<<<

Here we start from the general two-soliton solution of the U(2) model \cite{L14} and specialize it as follows. We choose $\eta_1=\eta_2=1$
(bound state at rest) and a diagonal matrix $\omega$ (no breathers). To get a real mean field $\Delta$, we have to pair the twist angles 
to $(\theta_1=\pi- \theta_2=\theta)$  and to choose $\omega_{11}=\omega_{22}$, as in the GN model. Moreover, it turns out that we need $\alpha_1=\alpha_2=\alpha$.
Consider the simplest case first, $\alpha=0$.
We find 
\begin{equation}
\Delta = \left( \begin{array}{cc} \frac{1+2V \cos 2 \theta + V^2  \cos^2 \theta}{1+ 2 V + V^2 \cos^2 \theta} & 0 \\ 0 & 1 \end{array}\right),  \quad V=e^{2 x \sin \theta}.
\label{8.1}
\end{equation}
This is nothing but the Dashen, Hasslacher, Neveu (DHN) baryon \cite{L18} in the upper component and the vacuum in the lower component. If $\theta$ is close to its maximum value of $\pi/2$,
the DHN scalar potential has the form of a well separated kink-antikink pair. Asymptotically, the vacuum is 1, but inside the baryon the vacuum -1 is approached.
In the present two-flavor case, the corresponding vacua are 1 (rotation) and $\tau_3$ (reflection).
A more transparent representation of the baryon mean field (\ref{8.1}) is 
\begin{eqnarray}
\Delta(\alpha=0) & = &  (1-g) - \tau_3 g ,
\nonumber \\
g & = &  \frac{2 V \sin^2 \theta}{1 + 2 V + V^2 \cos^2 \theta} ,
\nonumber \\
S_{\rm DHN} & = &   1-2 g .
\label{8.2}
\end{eqnarray}
The function $g$ vanishes asymptotically on both sides, in contrast to the kinklike function $f$ introduced above for a single kink.
We can now perform a rotation, leaving the asymptotic vacua unchanged but transforming the central vacuum into an
arbitrary reflection matrix. This yields the baryon solution for any $\alpha$,
\begin{equation}
\Delta(\alpha)  = 1-g -I(2\alpha) g.
\label{8.3}
\end{equation}
The same result would have been obtained by setting $\alpha_1=\alpha_2 =\alpha$ rather than 0 from the beginning. 
The maximum of $g$ is at
\begin{equation}
x_{\rm max} = - \frac{\ln (\cos \theta)}{2 \sin \theta}, \quad  g_{\rm max} = 1-\cos \theta.
\label{8.4}
\end{equation}
Hence, at the center of the baryon,
\begin{equation}
\Delta = \cos \theta - (1-\cos \theta)I(2 \alpha).
\label{8.5}
\end{equation}
If $\theta \approx \pi/2$ as appropriate for well separated kink and antikink, $\cos \theta \approx 0$ and
we see that we are in the reflection phase inside the baryon. 
Since the unitary transformation may be viewed as a change of frame, the angle $\alpha$ is irrelevant for the single
baryon. However it will become relevant in problems involving more than two 
kinks. The single particle spectrum of the O(2) baryon is the same as that of the DHN baryon in the GN model.
There are two bound states at energy $\pm \cos \theta$ and the spectrum is symmetric about 0.
Also the occupation fractions match those of the DHN baryon. 

To get a breather, we have to repeat the above calculation with an off-diagonal $\omega$ matrix.
We choose 
\begin{equation}
\omega = \left( \begin{array}{cc} \cosh \chi & \sinh \chi \\ \sinh \chi & \cosh \chi \end{array} \right), \quad {\rm det\,} \omega = 1.
\label{8.6}
\end{equation}
The results for the breather have the same general form as for the baryon, except that $g$ acquires a time dependence
in the rest frame,
\begin{equation}
g = \frac{2 V \sin \theta \left[ \cosh \chi \sin \theta - \sinh \chi \sin(2t  \cos \theta + \theta) \right] }{1 +2V\left[ \cosh \chi - \sin \theta \sinh \chi \sin(2t  \cos \theta + \theta) \right] + V^2 \cos^2 \theta}.
\label{8.7}
\end{equation}
At $\chi=0$ one recovers the static baryon, Eq.~(\ref{8.2}). 

So far, we have essentially reproduced the results of DHN \cite{L18} in a new setting. Now we can also look at more complicated multibaryon and
breather problems where new phenomena are expected. All we have to do is perform the calculation in the 
U(2) GN model and choose the parameters such that all mean fields are real. Judging from the experience with the one flavor GN and NJL models,
this should be more efficient than trying to determine solutions of the O(2) model directly. Up to now, we have only studied three-kink scattering
and baryon-kink scattering in the O(2) GN model along these lines. The resulting expressions are too lengthy to be written down here. We have found no
indication that the method does not work for more complicated many-particle collisions. To definitely confirm the quantum
integrability of the O(2) model, it would be nice if one could find closed analytical expressions for the scattering of any number of 
bound states, as in the one-flavor GN and NJL models, but this has to be left for future work. 

%<<<<<<<<<<<<<<<<<<<<<<<<<<<<<<<<<<<<<<<<<<<<<<<<<<<<<<<<<<<<<<<<<<<<<<<<<<<<<<<<<<<<<<<<<<<< <<<<<<<<<<<<<<<<<<<<<<<<<<<<<
%<<<<<<<<<<<<<<<<<<<<<<<<<<<<<<<<<<<<<<<<<<<<<<<<<<<<<<<<<<<<<<<<<<<<<<<<<<<<<<<<<<<<<<<<<<<<<<<<<<<<<<<<<<<<<<<<<<<<<<<<<<
\section{Massless hadrons, chiral spiral and phase diagram} 
\label{sect9}
%<<<<<<<<<<<<<<<<<<<<<<<<<<<<<<<<<<<<<<<<<<<<<<<<<<<<<<<<<<<<<<<<<<<<<<<<<<<<<<<<<<<<<<<<<<<<<<<<<<<<<<<<<<<<<<<<<<<<<<<<<<
%<<<<<<<<<<<<<<<<<<<<<<<<<<<<<<<<<<<<<<<<<<<<<<<

In this section, we briefly consider topics that have come up previously in the NJL model, the SU(2)$_L \times$SU(2)$_R$ isoNJL model
and the U(2)$_L \times$U(2)$_R$ GN model. Whenever such a model possesses a ``chiral circle" and massless mesons, the
possibility arises to generate both massless multifermion bound states and ``chiral spiral" type matter phases \cite{L15,L20}. Whereas the massless mesons
are related to small fluctuations around the vacuum into some flat direction (Goldstone modes), massless baryons correspond to one full turn around 
the chiral circle. The axial anomaly links winding number to fermion density. Because of the common charge conjugation symmetry, the situation
of the O(2) chiral GN model is perhaps closest to the one of the isoNJL model. We refer to Sec.~V of Ref.~\cite{L21} for the pertinent discussion
in the isoNJL model. Here, only the condensates $\bar{\psi}\psi$ and $\bar{\psi} i \gamma_5 \tau_3\psi $ had to be used. The crucial ingredient 
was the unitary transformation
\begin{equation}
\psi' = U\psi = e^{-ibx \gamma_5 \tau_3} \psi.
\label{9.1}
\end{equation}
If applied to the vacuum HF equation with potential $\gamma^0 m$, it generates an isospin chemical potential $b \tau_3$ from the kinetic term
and changes the mass term into the characteristic chiral spiral condensate
\begin{equation}
 U^{\dagger} (-i \gamma_5  \partial_x) U = - i  \gamma_5 \partial_x - b \tau_3, \quad U^{\dagger}\gamma^0 U = \gamma^0 \cos 2bx
+ i \gamma^1 \tau_3 \sin 2bx.
\label{9.2}
\end{equation}
This construction cannot be used for generating fermion density and the corresponding chemical potential in the two-flavor model. Here, 
the GN model with discrete chiral symmetry has taught us how to minimize the energy, namely by generating a real kink-antikink crystal described
by cnoidal functions. Referring to the literature for the details \cite{L22}, let us denote the self-consistent scalar potential by $S_{\rm GN}(x)$, with
temperature and density dependent shape. If one takes into account both fermionic and isospin chemical potentials, the HF potential in the isoNJL model 
assumes the product form
\begin{equation}
\Delta(x) = S_{\rm GN}(x) e^{-ibx \gamma_5 \tau_3}.
\label{9.3}
\end{equation}
The axial isospin chemical potential conjugate to the density $\psi^{\dagger} \gamma_5 \tau_3 \psi$ can also be invoked if one is interested
in chirally imbalanced states, but does not affect the mean field. The phase diagram of the isoNJL model in ($\mu,\mu_3,T$) space following from this scenario
consists of the GN phase diagram in the ($\mu,T$) plane translated rigidly into the $\mu_3$ direction; see \cite{L21}.

What does this teach us about the O(2) chiral GN model? In the above sketched results for the isoNJL model, only the condensates $\bar{\psi}\psi$
and $\bar{\psi}i \gamma_5 \tau_3 \psi$ have played a role. The choice of the three-direction in isospin is arbitrary and only used for convenience, since $\tau_3$ is diagonal.
One could equally well have used the two-direction. But then we would be in the same situation as in the O(2) case with a rotation vacuum and a chiral circle in the ($S_0,P_2$) plane. 
There is a one-to-one correspondence between SU(2) and O(2) chiral GN models, as far as these particular aspects are concerned.
Thus we can borrow the results \cite{L21} from the isoNJL model directly and get massless bound states and the whole phase diagram of the O(2) model almost
for free.

What would happen, had we started from the reflection vacuum in the ($S_1,S_3$) plane instead of the rotation vacuum? Here the transition to the isoNJL model
is less straightforward, but we certainly expect an equivalent picture. To induce rotation around the vacuum circle in the ($S_1,S_3$) plane now requires the
vector transformation
\begin{equation}
\psi' = U \psi = e^{-ibx \tau_2} \psi,
\label{9.4}
\end{equation}
without $\gamma_5$ in the exponent. This yields an axial isovector chemical potential and a chiral spiral mean field in the ($S_1,S_3$) reflection plane,
\begin{equation}
U^{\dagger}(-i \gamma_5  \partial_x) U = -i \gamma_5 \partial_x - b \gamma_5 \tau_2, \quad U^{\dagger} \gamma^0 \tau_3  U = \gamma^0 \left( \tau_3  \cos 2bx
+ \tau_1  \sin 2bx  \right).
\label{9.5}
\end{equation}
In this case, it is the axial isospin density that induces the inhomogeneous chiral spiral structure. This change of vector into axial chemical potentials
is also known from other dualities \cite{L8}, so that everything fits nicely together.

%<<<<<<<<<<<<<<<<<<<<<<<<<<<<<<<<<<<<<<<<<<<<<<<<<<<<<<<<<<<<<<<<<<<<<<<<<<<<<<<<<<<<<<<<<<<< <<<<<<<<<<<<<<<<<<<<<<<<<<<<<
%<<<<<<<<<<<<<<<<<<<<<<<<<<<<<<<<<<<<<<<<<<<<<<<<<<<<<<<<<<<<<<<<<<<<<<<<<<<<<<<<<<<<<<<<<<<<<<<<<<<<<<<<<<<<<<<<<<<<<<<<<<
\section{Duality between O(2)$_L\times$O(2)$_R$ GN model and ${\bf p}$GN model} 
\label{sect10}
%<<<<<<<<<<<<<<<<<<<<<<<<<<<<<<<<<<<<<<<<<<<<<<<<<<<<<<<<<<<<<<<<<<<<<<<<<<<<<<<<<<<<<<<<<<<<<<<<<<<<<<<<<<<<<<<<<<<<<<<<<<
%<<<<<<<<<<<<<<<<<<<<<<<<<<<<<<<<<<<<<<<<<<<<<<<

We label the fermion fields of the O(2) chiral GN model (\ref{3.3}) as $\psi_{k \ell}^{(i)}$ with $k$ the chirality ($1=L,2=R$), $\ell$ the flavor index (1,2) and
$i$ the color index ($i=1,...,N)$. Each Dirac field can be decomposed into two Majorana fields, 
\begin{equation}
\left( \begin{array}{c} \psi_{11}^{(i)}  \\ \psi_{12}^{(i)} \\ \psi_{21}^{(i)} \\ \psi_{22}^{(i)} \end{array} \right) = \frac{1}{\sqrt{2}} \left( \begin{array}{c}
\chi_1^{(i)} - i \chi_1^{(N+i)} \\ \chi_3^{(i)} - i \chi_3^{(N+i)} \\ \chi_2^{(N+i)} + i \chi_2^{(i)} \\ \chi_4^{(N+i)} + i \chi_4^{(i)} \end{array} \right).
\label{10.1}
\end{equation}
The labeling of the Majorana spinors has been chosen for later convenience. Here we only note that the even subscripts belong to right-handed and the odd subscripts to
left-handed Majorana spinors satisfying the anticommutation relations
\begin{equation}
\left\{ \chi_n^{(i)}(x),\chi_m^{(j)}(y) \right\} = \delta_{ij} \delta_{nm} \delta(x-y).
\label{10.2}
\end{equation}
The terms in the interaction Lagrangian can be regrouped as follows,
\begin{eqnarray}
{\cal L}_{\rm int} & = & \frac{g^2}{4} \left[ (\bar{\psi} \psi)^2 + (\bar{\psi} \tau_1 \psi)^2 + (\bar{\psi} \tau_3 \psi)^2 + (\bar{\psi} i \gamma_5 \tau_2 \psi)^2 \right]
\nonumber \\
& = & \frac{g^2}{8} \left[  (\bar{\psi}(1+\tau_3)\psi)^2 +  (\bar{\psi}(1-\tau_3)\psi)^2 + (\bar{\psi} (\tau_1 + i \gamma_5 \tau_2)\psi)^2 +  (\bar{\psi} (\tau_1 - i \gamma_5 \tau_2)\psi)^2 \right].
\label{10.3}
\end{eqnarray}
The motivation for the last line can be seen once we express everything in terms of Majorana spinors, 
\begin{equation}
\sum_{i=1}^N \bar{\psi}^{(i)} (1+ \tau_3 ) \psi^{(i)}  =  2i \sum_{i=1}^{N}\left( \chi_1^{(i)} \chi_2^{(i)} + \chi_1^{(N+i)} \chi_2^{(N+i)} \right) =  2i \sum_{i=1}^{2N}\left( \chi_1^{(i)} \chi_2^{(i)} \right) ,
\label{10.4}
\end{equation}
and similarly
\begin{eqnarray}
\sum_{i=1}^N \bar{\psi}^{(i)} (1- \tau_3 ) \psi^{(i)}  =   2i \sum_{i=1}^{2N} \left( \chi_3^{(i)} \chi_4^{(i)} \right) ,
\nonumber \\
\sum_{i=1}^N \bar{\psi}^{(i)} (\tau_1+ i \gamma_5 \tau_2 ) \psi^{(i)}  =   2i \sum_{i=1}^{2N}\left( \chi_1^{(i)} \chi_4^{(i)} \right) ,
\nonumber \\
\sum_{i=1}^N \bar{\psi}^{(i)} (\tau_1- i \gamma_5 \tau_2 ) \psi^{(i)}  =   2i \sum_{i=1}^{2N}\left( \chi_3^{(i)} \chi_2^{(i)} \right) .
\label{10.5}
\end{eqnarray}
The Lagrangian of the O(2) chiral GN model is thus turned into
\begin{eqnarray}
{\cal L}_{\rm O(2)} & = & -i \chi_1 \partial \chi_1 - i \chi_3 \partial \chi_3 + i \chi_2 \bar{\partial} \chi_2 + i \chi_4 \bar{\partial} \chi_4 
\nonumber \\
& &  - \frac{g^2}{2}  \left[ (\chi_1 \chi_2)^2 + (\chi_3 \chi_4)^2 + (\chi_1 \chi_4)^2 + (\chi_3 \chi_2)^2 \right],
\label{10.6}
\end{eqnarray}
with an implicit summation over $2N$ colors ($N$ Dirac fields are equivalent to 2$N$ Majorana fields).
In contrast to the original form in Eq.~(\ref{10.3}), expression (\ref{10.6}) is manifestly invariant under O(2)$_L \times$O(2)$_R$ since the vector with components ($\chi_1,\chi_3$) transforms 
under O(2)$_L$, the vector with components ($\chi_2,\chi_4$) under O(2)$_R$.

Consider now the pGN  model, Eq.~(\ref{1.7}),  obtained by ``self-dualizing" the NJL model.
As pointed out in the Introduction, the Lagrangian coincides with that of the classical  ZM model originally written in the form (\ref{1.4}).
Introducing Majorana spinors
\begin{equation}
\left( \begin{array}{c} \psi_1^{(i)} \\ \psi_2^{(i)} \end{array} \right) = \frac{1}{\sqrt{2}} \left( \begin{array}{c}  \chi_1^{(i)} - i \chi_3^{(i)} \\ \chi_4^{(i)} + i \chi_2^{(i)} \end{array} \right), \quad (i=1,..,N),
\label{10.7}
\end{equation}
it has already been shown in Ref.~\cite{L9} that the Lagrangian becomes
\begin{eqnarray}
{\cal L}_{\rm pGN} & = & -i \chi_1 \partial \chi_1 - i \chi_3 \partial \chi_3 + i \chi_2 \bar{\partial} \chi_2 + i \chi_4 \bar{\partial} \chi_4 
\nonumber  \\
& & - g^2 \left[ (\chi_1 \chi_2)^2 + (\chi_3 \chi_4)^2 +(\chi_1 \chi_4)^2 + (\chi_3 \chi_2)^2 \right].
\label{10.8}
\end{eqnarray}
Remarkably, this expression agrees with (\ref{10.6}). The coupling constants differ by a factor of 2, but this is compensated by the number of colors, $N$ in (\ref{10.8}) instead of $2N$
in (\ref{10.6}). Hence the 
O(2)$_L \times$O(2)$_R$ symmetric two-flavor GN model with $N$ colors is dual to the one-flavor pGN model with $2N$ colors. Since we have derived the O(2) 
symmetric model from the U(2) NJL model which has already
been solved in the large $N$ limit, we can now easily infer the solution of the dual model, not yet available in \cite{L9}. The solution
of many aspects of the O(2) GN model has already been discussed in the preceding sections. All we have to do is to reinterpret everything in the
dual language. This is the topic of the following section.

By eliminating the Majorana spinors from (\ref{10.1}) and (\ref{10.7}), we can express the Dirac fields of the O(2) GN model through
Dirac fields of the pGN model or vice versa,
\begin{equation}
\left( \begin{array}{c} \psi_{11}^{(i)} \\ \psi_{12}^{(i)} \\ \psi_{21}^{(i)} \\ \psi_{22}^{(i)} \end{array} \right) = \frac{1}{2}\left( \begin{array}{c} \psi_1^{(i)} + \psi_1^{(i)*}
- i \psi_1^{(N+i)} - i \psi_1^{(N+i)*} \\ \psi_1^{(N+i)} - \psi_1^{(N+i)*} + i \psi_1^{(i)} - i \psi_1^{(i)*} \\  \psi_2^{(i)} - \psi_2^{(i)*} - i \psi_2^{(N+i)}+ i \psi_2^{(N+i)*} \\
\psi_2^{(N+i)}  + \psi_2^{(N+i)*}  + i \psi_2^{(i)} + i \psi_2^{(i)*}\end{array} \right).
\label{10.9}
\end{equation}
As one can easily check, this is a Bogoliubov transformation at the level of Dirac spinors. It would have been difficult to find this transformation 
without introducing Majorana spinors at an intermediate step, but we can now discuss both sides of the duality in the more familiar Dirac language.
In particular, using (\ref{10.9}), we can express the relevant bilinears of the two-flavor model through bilinears of the one-flavor model as follows 
\begin{eqnarray}
\bar{\psi}^{(i)}\psi^{(i)} & = & \psi_1^{(i)*} \psi_2^{(i)} + \psi_2^{(i)*} \psi_1^{(i)} + \psi_1^{(N+i)*} \psi_2^{(N+i)} + \psi_2^{(N+i)*} \psi_1^{(N+i)} ,
\nonumber \\
\bar{\psi}^{(i)} \tau_1 \psi^{(i)} & = & i\left( \psi_1^{(i)*} \psi_2^{(i)*} - \psi_2^{(i)}\psi_1^{(i)} +  \psi_1^{(N+i)*} \psi_2^{(N+i)*} - \psi_2^{(N+i)}\psi_1^{(N+i)} \right),
\nonumber \\
\bar{\psi}^{(i)} \tau_3 \psi^{(i)} & = & \psi_1^{(i)} \psi_2^{(i)} + \psi_2^{(i)*} \psi_1^{(i)*} + \psi_1^{(N+i)} \psi_2^{(N+i)} + \psi_2^{(N+i)*} \psi_1^{(N+i)*},
\nonumber \\
\bar{\psi}^{(i)}i \gamma_5 \tau_2 \psi^{(i)} & = & i \left(  \psi_1^{(i)*} \psi_2^{(i)}  - \psi_2^{(i)*} \psi_1^{(i)} + \psi_1^{(N+i)*} \psi_2^{(N+i)}  - \psi_2^{(N+i)*} \psi_1^{(N+i)}\right).
\label{10.10}
\end{eqnarray}
The left-hand side refers to the O(2) GN model, the right-hand side to the pGN model, and the equations hold for each color index $i=1,...,N$. The first line 
shows that the scalar condensate has the same meaning on both sides of the duality. The $i\gamma_5 \tau_2$ condensate of the O(2) model 
corresponds to the pseudoscalar condensate of the pGN model. The $\tau_3$ and $\tau_1$ condensates of the O(2) model go over into the two kinds of  
Cooper pair condensates.

Important bilinear observables not present in the O(2) Lagrangian are the isospin vector and axial vector densities (only the $\tau_2$ components belong to a conserved
current),
\begin{eqnarray}
\psi^{(i)\dagger}\tau_2  \psi^{(i)} & = & \psi_1^{(i)*} \psi_1^{(i)} + \psi_2^{(i)*} \psi_2^{(i)} + \psi_1^{(N+i)*} \psi_1^{(N+i)} + \psi_2^{(N+i)*} \psi_2^{(N+i)},
\nonumber \\
\psi^{(i)\dagger}\gamma_5 \tau_2  \psi^{(i)} & = & - \psi_1^{(i)*} \psi_1^{(i)}+ \psi_2^{(i)*} \psi_2^{(i)} - \psi_1^{(N+i)*} \psi_1^{(N+i)} + \psi_2^{(N+i)*} \psi_2^{(N+i)}.
\label{10.12}
\end{eqnarray}
Thus the $\tau_2$ isospin density corresponds to the fermion density $\psi^{\dagger}\psi$ of the pGN model, the $\gamma_5 \tau_2$ axial isospin density to the fermion axial density
$\psi^{\dagger} \gamma_5 \psi$. 

Note that all observables discussed so far are color singlets on both sides of the duality. It is interesting to understand what happens
to the U(1) vector symmetry and fermion density of the O(2) model upon dualization. If we translate the U(1) transformation $\psi_{k\ell}^{(i)} \to e^{i \alpha} \psi_{k\ell}^{(i)}$
into the dual language, we find that it reduces to the following orthogonal, color dependent transformation
\begin{eqnarray}
\left( \begin{array}{c} \psi_1^{(i)} \\ \psi_1^{(N+i)} \end{array} \right) & \to & \left( \begin{array}{rr} \cos \alpha & \sin \alpha \\ -\sin \alpha & \cos \alpha \end{array} \right)
\left( \begin{array}{c} \psi_1^{(i)} \\ \psi_1^{(N+i)} \end{array} \right),
\nonumber \\
\left( \begin{array}{c} \psi_2^{(i)} \\ \psi_2^{(N+i)} \end{array} \right) & \to & \left( \begin{array}{rr} \cos \alpha & \sin \alpha \\ -\sin \alpha & \cos \alpha \end{array} \right)
\left( \begin{array}{c} \psi_2^{(i)} \\ \psi_2^{(N+i)} \end{array} \right).
\label{10.13}
\end{eqnarray}
This is a subgroup of the original O(2$N$) symmetry of the pGN model with 2$N$ colors. 
Consequently, the 
conserved current in the dual pGN model is not a color singlet, but has the form
\begin{eqnarray}
\rho_R & \to & i \sum_{i=1}^N \left( \psi_2^{(N+i)*} \psi_2^{(i)} - \psi_2^{(i)*} \psi_2^{(N+i)} \right),
\nonumber \\
\rho_L & \to & i \sum_{i=1}^N \left( \psi_1^{(N+i)*} \psi_1^{(i)} - \psi_1^{(i)*} \psi_1^{(N+i)} \right),
\nonumber \\
\rho & = & \rho_R+ \rho_L , \quad j = \rho_5 = \rho_R-\rho_L.
\label{10.15}
\end{eqnarray}
Since we are not dealing with a gauge theory and color confinement, we see nothing wrong with this color dependence.

Finally, let us look at yet another self-dual model that has been discussed in Ref.~\cite{L9}. By self-dualizing the GN model with discrete chiral symmetry
rather than the NJL model, one gets the Lagrangian of the self-dual GN (sdGN) model
\begin{equation}
{\cal L}_{\rm sdGN} =   -2i \psi_1^{(i)*}\partial \psi_1^{(i)} + 2 i \psi_2^{(i)*} \bar{\partial} \psi_2^{(i)}  + \frac{g^2}{2}\left[ (  \psi_1^{(i)*} \psi_2^{(i)} + \psi_2^{(i)*} \psi_1^{(i)} )^2
+  (  \psi_1^{(i)*} \psi_2^{(i)*} + \psi_2^{(i)} \psi_1^{(i)} )^2\right].
\label{10.16}
\end{equation}
In \cite{L9} it came as a surprise that this is again equivalent to two independent GN models. If we apply the strategy developed in the present section to the sdGN model
and first translate it into Majorana spinors using (\ref{10.7}), we find 
\begin{equation}
{\cal L}_{\rm sdGN} =  -i \chi_1 \partial \chi_1 - i \chi_3 \partial \chi_3 + i \chi_2 \bar{\partial} \chi_2 + i \chi_4 \bar{\partial} \chi_4 
 - g^2 \left[ (\chi_1 \chi_2)^2 + (\chi_3 \chi_4)^2 \right].
\label{10.17}
\end{equation}
Upon using the ``dictionary" (\ref{10.4}) and (\ref{10.5}), this is equivalent to a U(2) model where only two out of the eight original interaction terms are kept,
\begin{equation}
{\cal L} = \bar{\psi} i \partial \!\!\!/ \psi + \frac{g^2}{4} \left[ (\bar{\psi}\psi)^2 + (\bar{\psi}\tau_3\psi)^2\right].
\label{10.18}
\end{equation}
Since the two isospin channels decouple, the fact that one gets two independent GN models is now trivial. This shows once more the advantage of 
using Majorana spinors as an intermediate step for discovering or elucidating dualities.

%<<<<<<<<<<<<<<<<<<<<<<<<<<<<<<<<<<<<<<<<<<<<<<<<<<<<<<<<<<<<<<<<<<<<<<<<<<<<<<<<<<<<<<<<<<<< <<<<<<<<<<<<<<<<<<<<<<<<<<<<<
%<<<<<<<<<<<<<<<<<<<<<<<<<<<<<<<<<<<<<<<<<<<<<<<<<<<<<<<<<<<<<<<<<<<<<<<<<<<<<<<<<<<<<<<<<<<<<<<<<<<<<<<<<<<<<<<<<<<<<<<<<<
\section{Summary: an integrable model with chiral symmetry breaking and Cooper pairing} 
\label{sect11}
%<<<<<<<<<<<<<<<<<<<<<<<<<<<<<<<<<<<<<<<<<<<<<<<<<<<<<<<<<<<<<<<<<<<<<<<<<<<<<<<<<<<<<<<<<<<<<<<<<<<<<<<<<<<<<<<<<<<<<<<<<<
%<<<<<<<<<<<<<<<<<<<<<<<<<<<<<<<<<<<<<<<<<<<<<<<

There has been quite some interest in four-fermion theories featuring a competition between particle-hole pairing and Cooper pairing \cite{L23,L24,L25,L26,L27,L28}, 
triggered partly by 
predictions of color superconductivity in quantum chromodynamics \cite{L29}. These works are dealing mostly with the phase diagrams of the models considered.
As we have shown in the present paper, the pGN model is an example where the coexistence of chiral symmetry breaking (CSB) and Cooper pairing arises in a highly symmetric fashion.
As a consequence, the resulting model is integrable and can be solved as completely as the GN or NJL models. This confirms our earlier speculation \cite{L9}
and extends the range of integrable field theory models in 1+1 dimensions into an interesting direction. Crucial for the new insights was a mapping between a
O(2)$_L \times$O(2)$_R$ symmetric GN model, which had to be constructed for this purpose, and the pGN model, as summarized in Table~\ref{tab2}.
The pGN model has been obtained by ``self-dualizing" the one-flavor NJL model with respect to the transformation (\ref{1.6}),
\begin{equation}
\psi = \left( \begin{array}{c} \psi_1 \\ \psi_2 \end{array} \right) \quad \longrightarrow \quad \psi_d = \left( \begin{array}{c} \psi_1^{\dagger} \\ \psi_2 \end{array} \right).
\label{11.1}
\end{equation}
Table~\ref{tab2} shows the correspondence between the condensates of the two equivalent models. Since the solution of the O(2) chiral GN model could be derived 
with the machinery developed for the U(2) chiral GN model, one can take over all the results collected above and reinterpret them in terms of the physics of 
relativistic superconductors. Recall that the Dirac fields and Majorana spinors have $N$ color components in the O(2) model but 2$N$ color components in the pGN model. The 
labeling of the Majorana spinor indices in Eqs.~(\ref{10.1}) and (\ref{10.7}) has been chosen with hindsight, so that the expressions in the last column of Table~\ref{tab2} 
are identical in both models. Likewise, the Lagrangians are indistinguishable if expressed in terms of Majorana spinors, see Eqs.~(\ref{10.6}) and (\ref{10.8}).
 
\begin{center}
\begin{table}
\begin{tabular}{|c|c|c|c|}
\hline
 condensate &  O(2)$_L \times$O(2)$_R$  &  pGN  & Majorana spinors    \\
\hline
$S_0$ &  $\bar{\psi}\psi$ & $\bar{\psi}\psi$ & $i (\chi_1 \chi_2+ \chi_3 \chi_4)$ \\
$P_2$ &  $ \bar{\psi}i \gamma_5 \tau_2 \psi $ & $  \bar{\psi} i \gamma_5 \psi $ & $ i (\chi_1 \chi_4 + \chi_2 \chi_3)$ \\
$S_3$ &  $\bar{\psi} \tau_3 \psi $ & $ \bar{\psi}_d \psi_d $  & $ i (\chi_1 \chi_2- \chi_3 \chi_4) $  \\
$S_1$ & $ \bar{\psi} \tau_1 \psi $ & $ \bar{\psi}_d i \gamma_5 \psi_d $  & $ i (\chi_1 \chi_4-\chi_2 \chi_3)$ \\
\hline
\end{tabular}
\caption{Correspondence between bilinears in the two dual models. The Majorana notation is common to both, but summation is over
$N$ colors (O(2) model) and 2$N$ colors (pGN model), respectively.}
\label{tab2}
\end{table}
\end{center}

Let us briefly review the results of the large $N$ O(2) model in the light of the pGN model.

\begin{itemize}
\item
Lagrangian: The Lagrangians can be cast into a form that emphasizes the correspondence, using Table~\ref{tab2} and the definition (\ref{11.1}),
\begin{eqnarray}
{\cal L}_{\rm O(2)} & = &   \bar{\psi} i \partial \!\!\!/ \psi + \frac{g^2}{4} \left[ (\bar{\psi} \psi)^2 + (\bar{\psi}i \gamma_5 \tau_2  \psi)^2 + (\bar{\psi}\tau_3  \psi)^2 + (\bar{\psi}\tau_1  \psi)^2\right],
\nonumber \\
{\cal L}_{\rm pGN} & = &   \bar{\psi} i \partial \!\!\!/ \psi + \frac{g^2}{4} \left[ (\bar{\psi} \psi)^2 + (\bar{\psi}i \gamma_5 \psi)^2 + (\bar{\psi}_d \psi_d)^2 + (\bar{\psi}_d i \gamma_5  \psi_d)^2\right].
\label{11.2}
\end{eqnarray}
The O(2)$_L \times$O(2)$_R$ chiral symmetry of the first model matches the Pauli-G\"ursey symmetry of the second one. 
\item 
Vacuum: The O(2) model has two possible vacuum circles referred to as rotation and reflection vacua above. In the pGN model, this corresponds to the CSB
vacuum (condensates $\bar{\psi}\psi,\bar{\psi} i \gamma_5 \psi$) or the Cooper paired vacuum (condensates $\bar{\psi}_d\psi_d,\bar{\psi}_d i \gamma_5 \psi_d$).
The two connected components of the O(2) group are exactly what it takes to describe these two distinct possibilities. The pGN model gives a more physical picture of
what it means to break chiral symmetry spontaneously if the symmetry group is continuous, but not connected.
\item 
Kink: The kink becomes a domain wall between superconducting and normal phase. There is no kink inside a single phase. One can study dynamical
problems such as scattering of any number of such domain walls in closed analytical form, describing time dependent configurations of CSB and Cooper pairing domains.
\item
DHN-type baryon: This bound state of two twisted kinks in the O(2) model is nothing but a configuration where one phase is separated from the other phase
by two walls. If the exterior phase is chosen as Cooper paired phase, then the interior phase is normal and we get a relativistic toy model for a Josephson 
junction. Dynamical problems including several such objects as well as single domain walls can be solved, as well as the interaction of fermions or Cooper pairs with domain walls.
Breathers can be interpreted as excited, periodically oscillating Josephson junctions. 
\item
Massless many-fermion states and phase diagram: In the pGN model, fermion density can be generated by a local chiral transformation, such as in the NJL
model. This makes the chiral spiral configuration optimal for fermionic matter, in the normal phase. In the Cooper pairing phase, the same mathematics
would give rise to an inhomogeneous LOFF phase of the superconductor \cite{L30,L31}. The two are just two ways of interpreting the same physics, in the pGN model.
The phase diagram with vector and axial vector fermion chemical potentials of the pGN model can be taken over from the NJL model. Fermion density
of the O(2) model becomes a color dependent density (\ref{10.15}). If one introduces the  conjugated chemical potential, the kink crystal of the
GN model must come into the picture and color O(2$N$) breaks down to O($N$). 
\end{itemize}

While it is easy to write down four-fermion models with both $pp$- and $ph$-pairing, it is not easy to find integrable ones. The only one that had
been found so far is the sdGN model \cite{L9}. Since this has turned out to be a trivial double copy of the standard GN model, it gives few new
insights. All other integrable models known so far had either CSB or Cooper pairing, but not both. In this sense, the quantum ZM or pGN model
is a novel and potentially useful member of the family of exactly solvable field theoretic models which deserves further studies.


\begin{thebibliography}{99}
\bibitem{L1}
V. E. Zakharov and A. V. Mikhailov, Commun. Math. Phys. {\bf 74}, 21 (1980).
\bibitem{L2}
D. J. Gross and A. Neveu, Phys. Rev. D {\bf 10}, 3235 (1974).
\bibitem{L3}
Y. Nambu and G. Jona-Lasinio, Phys. Rev.  {\bf 124}, 246 (1961).
\bibitem{L4} 
D. A. Takahashi and M. Nitta, Phys. Rev. Lett. {\bf 110}, 131601 (2013).
\bibitem{L5}
G. V. Dunne and M. Thies, Phys. Rev. Lett. {\bf 111}, 121602 (2013).
\bibitem{L6}
G. V. Dunne and M. Thies, Phys. Rev. D {\bf 89}, 025008 (2014).
\bibitem{L7}
A. Chodos, H. Minakata, and F. Cooper, Phys. Lett. B {\bf 449}, 260 (1999).
\bibitem{L8}
M. Thies, Phys. Rev. D {\bf 68}, 047703 (2003).
\bibitem{L9}
M. Thies, Phys. Rev. D {\bf 90}, 105017 (2014).
\bibitem{L10} 
W. Pauli, Nuovo Cimento {\bf 6}, 204 (1957).
\bibitem{L11}
F. G\"ursey, Nuovo Cimento {\bf 7}, 411 (1958).
\bibitem{L12}
J. Milanovic, {\em Das perfekte Gross-Neveu Modell}, Diplomarbeit (Universit\"at
Erlangen-N\"urnberg, Erlangen, 2004). 
\bibitem{L13}
D. A. Takahashi, Phys. Rev. B {\bf 93}, 024512 (2016).
\bibitem{L14}
M. Thies, J. Phys. A {\bf 55}, 015401 (2022).
\bibitem{L15}
L. L. Salcedo, S. Levit, and J. W. Negele, Nucl. Phys. {\bf B361}, 585 (1991).
\bibitem{L16}
R. Pausch, M. Thies, and V. L. Dolman, Z. Phys. A {\bf 338}, 441 (1991).
\bibitem{L17}
S.-S. Shei, Phys. Rev. D {\bf 14}, 535 (1976).
\bibitem{L18}
R. F. Dashen, B. Hasslacher, and A. Neveu, Phys. Rev. D {\bf12}, 2443 (1975).
\bibitem{L19}
J. Feinberg, Phys. Rev. D {\bf 51}, 4503 (1995).
\bibitem{L20}
V. Sch\"on and M. Thies, Phys. Rev. D {\bf 62}, 096002 (2000).
\bibitem{L21}
M. Thies, Phys. Rev. D {\bf 93}, 085024 (2016).
\bibitem{L22}
O. Schnetz, M. Thies, and K. Urlichs, Ann. Phys. (Amsterdam) {\bf 314}, 425 (2004).
\bibitem{L23}
N. Ilieva and W. Thirring, Nucl. Phys. {\bf B565}, 629 (2000).
\bibitem{L24}
A. Chodos, F. Cooper, W. Mao, H. Minakata, and A. Singh, Phys. Rev. D {\bf 61}, 045011 (2000).
\bibitem{L25}
K. G. Klimenko, R. N. Zhukov, and V. Ch. Zhukovsky, Phys. Rev. D {\bf 86}, 105010 (2012).
\bibitem{L26}
D. Ebert, T. G. Khunjua, K. G. Klimenko, and V. Ch. Zhukovsky, Int. J. Phys. A {\bf 29}, 1450025 (2014).
\bibitem{L27}
D. Ebert, T. G. Khunjua, K. G. Klimenko, and V. Ch. Zhukovsky, Phys. Rev. D {\bf 91}, 105024 (2015).
\bibitem{L28}
D. Ebert, T. G. Khunjua, K. G. Klimenko, and V. C. Zhukovsky, Phys. Rev. D {\bf 93}, 105022 (2016).
\bibitem{L29}
K. Rajagopal and F. Wilczek, in {\em At the Frontier of Particle Physics: Handbook of QCD}, Boris Ioffe Festschrift, edited by
M. Shifman (World Scientific, Singapore, 2001), Vol. 3, Ch. 35, p. 2061.
\bibitem{L30}
P. Fulde and R. A. Ferrell, Phys. Rev. {\bf 135}, A550 (1964).
\bibitem{L31}
A. I. Larkin and Yu. N. Ovchinnikov, Zh. Eksp. Teor. Fiz. {\bf 47}, 1136 (1964) and
Sov. Phys. JETP {\bf 20}, 762 (1965).
\end{thebibliography}
\end{document}